\documentclass[prl,aps,twocolumn,showpacs]{revtex4}
\usepackage{epsfig}
\usepackage{graphicx}
\usepackage{textcomp}

\begin{document}
\title{Extinction of an infectious disease: a large fluctuation in a non-equilibrium system}
\author{Alex Kamenev$^1$ and Baruch Meerson$^{1,2}$}
\affiliation{$^1$Department of Physics, University of Minnesota,
Minneapolis, Minnesota 55455, USA\\$^2$Racah Institute of Physics,
Hebrew University of Jerusalem, Jerusalem 91904, Israel}


\begin{abstract}
We develop a theory of first passage processes in stochastic non-equilibrium
systems of 
birth-death type using two closely related epidemiological models as examples.
Our method employs the probability generating
function technique in conjunction with the eikonal approximation. In this
way the problem is reduced to finding the optimal path to extinction: a heteroclinic  trajectory of an effective multi-dimensional classical Hamiltonian system. We compute this trajectory and mean extinction time of the disease numerically and uncover a non-monotone, spiral path to extinction of a disease. We also obtain analytical results  close to a bifurcation point, where the problem is described by a Hamiltonian previously identified in one-species population models.

\end{abstract}

\maketitle

Statistics of large fluctuations in
stochastic non-equilibrium systems has received much
recent attention \cite{large_fluctuations}. While the equilibrium
fluctuation probability
is determined by the Boltzmann distribution,
there is no similar general principle away from
equilibrium. The underlying reason is the absence of time
reversal symmetry between the relaxation and excitation dynamics in
out-of-equilibrium systems. As a
consequence, the most probable fluctuation path is \textit{not}
determined by the relaxation trajectory of the underlying
deterministic system.

An
important class of stochastic non-equilibrium systems
is reaction kinetics, or birth-death, systems \cite{Gardiner}. Rather than
being caused by external factors, the noise
in these systems is intrinsic,  as it originates
from discreteness of the reacting agents and random character of their interactions.
When the typical number of agents is large,
the Fokker-Planck (FP) approximation to the master equation, see \textit{e.g.} Ref. \cite{Gardiner},
can accurately describe small deviations from the probability distribution maxima. It
fails, however, in determining the probability of
large fluctuations \cite{Gaveau,Kamenev1,Assaf}. Therefore,
developing adequate theoretical tools for dealing with
large fluctuations is an important task.

One of the areas where the birth-death models have been very successful
is mathematical epidemiology, see Refs.
\cite{books}. In this Letter we
consider two closely related models of spread of disease in a population.
Although they have served as standard multi-population epidemiological models, the analysis of large fluctuations in each of these models has not been satisfactory. We will use the two models as prototypical
examples of multi-dimensional  stochastic non-equilibrium
systems.

Observing the dynamics of a disease in a finite population, one notices the remarkable phenomenon of extinction
of the disease in a finite time. The expected time to extinction
(and the possibility to affect it) is of great
practical interest. Here we develop an efficient theoretical
approach capable of computing, among other things, this quantity. The approach employs  the probability generating function formalism in conjunction with the eikonal approximation. In this way the problem is reduced to the dynamics of an effective classical Hamiltonian system.
The intrinsic-noise-induced extinction of the disease proceeds, with a high probability,
along the optimal path:
a special (heteroclinic) trajectory in the phase space of the classical
Hamiltonian flow. An additional challenge of this type of problems is in the fact that the emerging multi-dimensional  Hamiltonian flows are generally non-integrable. We
compute the optimal path, and the mean extinction time of the disease,
numerically and also obtain analytical results close to a bifurcation point.

\textit{Model.} Let us consider two stochastic epidemiological models: the endemic SI model and the endemic SIR model. In the SI model the host population is divided into two dynamic sub-populations: Susceptible (S) and Infected (I).  The model is specified by the set of reactions and their rates given in Table 1. We can always represent the renewal rate (an independent parameter of the model) as $\mu N$, where $N$ scales as the total population size in a steady state. Taking $\mu_I>\mu$, one allows an increased death rate of the infected.

\begin{table}[ht]
\begin{ruledtabular}
\begin{tabular}{|c|c|c|}
 Event & Type of transition &  Rate\\
  \hline
  Infection & $S\to S-1, \, I\to I+1$ &  $(\beta/N) SI$\\
  Renewal of susceptible & $S\to S+1$ & $\mu N$ \\
  Death of susceptible & $S\to S-1$ & $\mu S$ \\
  Death of infected & $I\to I-1$ & $\mu_I I$ \\
\end{tabular}
\end{ruledtabular}
\caption{Transition rates for the stochastic SI model}\label{table}
\end{table}

The endemic SIR model deals, in addition to the $S$- and $I$-sub-populations, with a third sub-population: Recovered (R), with the recovery rate $\gamma I$. It is assumed that the recovered cannot become susceptible. The death rate of the recovered is $\mu_R R$. The endemic SIR model (which generalizes the original SIR model: the one without renewal and death) gives a satisfactory description to the spread of measles, mumps and rubella \cite{books}.

Let us briefly review the deterministic, or mean-field version of the SIR model:
\begin{eqnarray}
  \dot{S} &=& \mu N-\mu S-(\beta/N) S\,I\,, \label{Sdot} \\
  \dot{I} &=&  -\mu_I  I -\gamma I + (\beta/N) S\,I \,,  \label{Idot} \\
  \dot{R} &=& -\mu_R R +\gamma I   \label{Rdot}\,.
\end{eqnarray}
As the dynamics of $S$ and $I$ decouples from that of $R$, the SIR model is effectively two-population, and we will not deal with  the $R$-dynamics.
Furthermore, one immediately notices that, by putting $\mu_I+\gamma=\Gamma$, the S- and I-dynamics in the SIR model becomes identical to that in the SI model, up to interchange of
$\mu_I$ and $\Gamma$. This also holds for the stochastic versions of the two models, and so we can treat them on equal footing, using $\Gamma$ for the effective death rate constant of infected.

For a sufficiently high infection rate,
$\beta > \Gamma$, there is an attracting fixed point
\begin{equation}\label{stablefp}
    \bar{S} = \frac{\Gamma}{\beta}\,N\,,\;\;\;\;\;\;\bar{I} = \frac{\mu (\beta-\Gamma)}{\beta\Gamma}\,N
\end{equation}
which describes an endemic infection level, and an unstable fixed point $\bar{S}=N, \,\bar{I}=0$ which describes an uninfected population. At $\mu < 4 \,(\beta-\Gamma) (\Gamma/\beta)^2$
the attracting fixed point is a stable focus,
while in the opposite case it is a stable node. The inverse of the real part of the eigenvalues (for the focus), or the inverse of the smaller of the eigenvalues (for the node) yields the characteristic relaxation time $\tau_r$ towards the ``endemic point".

The stochastic formulation of the SI and SIR models accounts for the demographic stochasticity and random character of contacts between the susceptible and infected. The master equation for the probability $P_{n,m}(t)$ of finding
$n$ susceptible and $m$ infected individuals has the form
\begin{eqnarray}
\label{master}
\dot{P}_{n,m}\!\!&=&\!\!\mu \left[N(P_{n-1,m}-P_{n,m})
+(n+1)P_{n+1,m}-n P_{n,m}\right]\nonumber\\
&+&\!\! \Gamma\left[(m+1)P_{n,m+1}-mP_{n,m}\right]\nonumber\\
&+&\!\! (\beta/N) \left[(n+1) (m-1)P_{n+1,m-1}-nm P_{n,m}\right]\,,
\end{eqnarray}
and the total population size is fluctuating in time.  We will be interested in the regime where the fluctuations are relatively weak. In this case, after the relaxation time $\tau_r$ a long-lived (quasi-stationary) distribution is formed that has a bi-variate gaussian peak with relative width
$\sim N^{-1/2}$ around the stable state (\ref{stablefp}) of the mean-field description \cite{Herwaarden,Nasell,cycles1}. The long-time behavior of the stochastic model is quite remarkable: due to a rare sequence of discrete events the disease goes extinct in a finite time. Given that a major outbreak of the disease occurred, what is the mean extinction time  $\tau$ of the disease? For the endemic SIR model this question was addressed previously \cite{Herwaarden,Nasell} in the framework of the
van Kampen system size expansion that brings about the approximate FP equation \cite{Gardiner}. Our approach considerably (exponentially) improves on these earlier results.  In the regime we are interested in $\tau$ is exponentially large compared with the relaxation time $\tau_r$.
The presence of the large parameter facilitates the use of the eikonal approximation: either directly in the master equation, as suggested by Dykman \textit{et al.} \cite{dykman1}, or in the evolution equation for the probability generating function, as suggested by Elgart and Kamenev \cite{Kamenev1}.

\textit{Probability generation function and eikonal approximation.}  We adopt the latter approach and introduce the probability generating function $G(p_S,p_I,t)=\sum_{n,m=0}^{\infty}p_S^{n}p_I^{m}P_{n,m}(t)$. Once $G(p_S,p_I,t)$ is found, the probabilities $P_{n,m}(t)$ are given by the coefficients of its Taylor expansion around $p_S=p_I=0$.  Using the master equation (\ref{master}), we obtain an evolution equation for $G$: $\partial_t G=\hat{H}G$ with the effective Hamiltonian operator
\begin{eqnarray}\label{hamiltonian}
\hat{H}&=&\ \mu (p_S-1)(N-\partial_{p_S})-\Gamma (p_I-1)\partial_{p_I}  \nonumber \\
&-&(\beta/N) (p_S -p_I)p_I\partial^2_{p_S p_I} \,.
\end{eqnarray}
In contrast to the FP equation this equation is exact \cite{spectral}.

The eikonal ansatz is $G(p_S,p_I,t)=\exp[-{\cal S}(p_S,p_I,t)]$, where
${\cal S}\gg 1$. Neglecting the second derivatives of ${\cal S}$ with respect to $p_S$ and $p_I$, we arrive at a Hamilton-Jacobi equation $\partial_t{\cal S}+H=0$ in the $p$-space
with the classical Hamiltonian $H(S,I,p_S,p_I)$:
\begin{equation}\label{hamiltonian1}
H= \mu (p_S-1)(N-S) - \Gamma  (p_I-1)I -(\beta/N) (p_S-p_I)p_I SI\,,
\end{equation}
where $S=-\partial_{p_S} {\cal S}$  and $I=-\partial_{p_I} {\cal S}$. The structure of four-dimensional (4D) phase space, defined by the Hamiltonian  (\ref{hamiltonian1}), provides a fascinating and instructive insight into the disease extinction dynamics. As $H$ does not depend explicitly on time, it is an integral of motion: $H(S,I,p_S,p_I)=E=const$. All the mean-field trajectories, described by Eqs.~(\ref{Sdot}) and (\ref{Idot}), lie in the zero-energy, $E=0$, two-dimensional plane $p_S=p_I=1$. The attracting fixed point (\ref{stablefp}) of the mean-field theory becomes
a hyperbolic point $A=[\bar{S},\bar{I},1,1]$ in the 4D phase space with two stable and two unstable eigenvalues (the sum of which is zero) and respective eigenvectors. There are two more zero-energy fixed points in the system: the point $C=[N,0,1,1]$ which is present in the mean-field description and the non-mean-field point $B=[N,0,1,\Gamma/\beta]$ which we call fluctuational. Both of them are hyperbolic and describe extinction of the disease. The presence of a \textit{fluctuational} fixed point, related to extinction, is characteristic of a class of stochastic birth-death systems \cite{Kamenev1,Herwaarden,Kamenev2,DykmanSIS}.

The most probable sequence of discrete events, bringing the system from the endemic state to extinction of the disease, is given by the optimal path that minimizes the action ${\cal S}$
\cite{dykman1,freidlin+graham}. The optimal path must be a zero-energy heteroclinic  trajectory. This trajectory exits, at $t=-\infty$, the ``endemic" point $A$  along its two-dimensional  unstable manifold and enters, at $t=\infty$, the fluctuational disease extinction point $B$, along its two-dimensional  stable manifold. As in one-dimensional birth-death systems \cite{Kamenev1,Kamenev2}, one can show that there is no trajectory going directly from $A$ to $C$.  Therefore, the fluctuational extinction point $B$, not present in the mean-field dynamics, plays a crucial role in the disease extinction.

Up to a pre-exponent, the mean extinction time of the disease is $\tau\propto\tau_r \exp({\cal S}_0)$ \cite{matkowsky}, where
\begin{equation}\label{action1}
    {\cal S}_0 = \int_{-\infty}^\infty (p_S \dot{S}+p_I \dot{I})\,dt\,,
\end{equation}
and the integration is performed along the (zero-energy) optimal path.  As in any generic multi-dimensional Hamiltonian system, the optimal path can be computed only numerically. In the following we present two typical examples of such computation, and also consider an important limit when the computation can be performed analytically, by exploiting time scale separation. First we introduce  new coordinates $x=S/N-1$ and $y=I/N$, time $\tilde{t}=\mu t$, momenta $p_{x,y}=p_{S,I}-1$ and  bifurcation parameter $\delta=1-\Gamma/\beta$, $0<\delta<1$. The action (\ref{action1}) can now be rewritten as
${\cal S}_0=N  \sigma $, where
$ \sigma(K,\delta)$ is the action along the optimal path, generated by the Hamiltonian
\begin{equation}\label{hamiltonian2}
\tilde{H}= -p_x x -K \left[(1-\delta) p_y+(p_x-p_y)(p_y+1)(x+1)\right]y
\end{equation}
and $K=\beta/\mu>1$. The fixed points $A$, $B$, and $C$ become
$$
\left[-\delta,\;\; \frac{\delta}{K(1-\delta)},0,0\right],\;\;[0,0,0,-\delta],\;\;\mbox{and}\;\;[0,0,0,0],
$$
respectively.

\begin{figure}[ht]
\begin{tabular}{cc}
\hspace{-6mm}
  \includegraphics[scale=0.50]{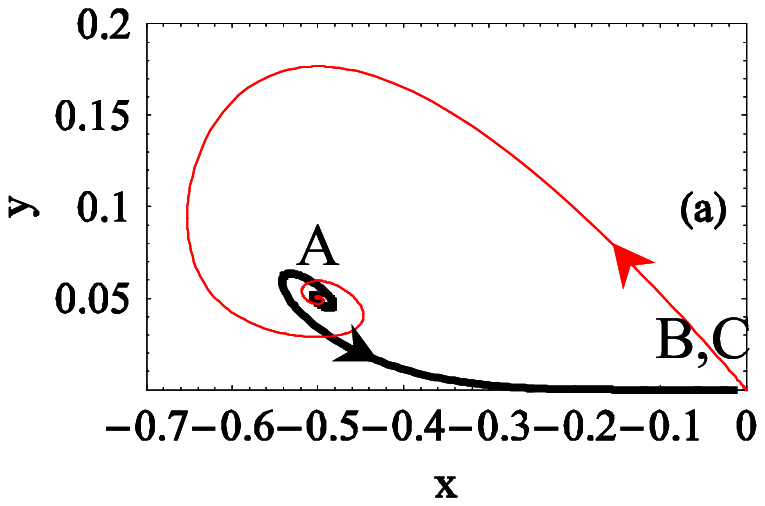} &
  \includegraphics[scale=0.45]{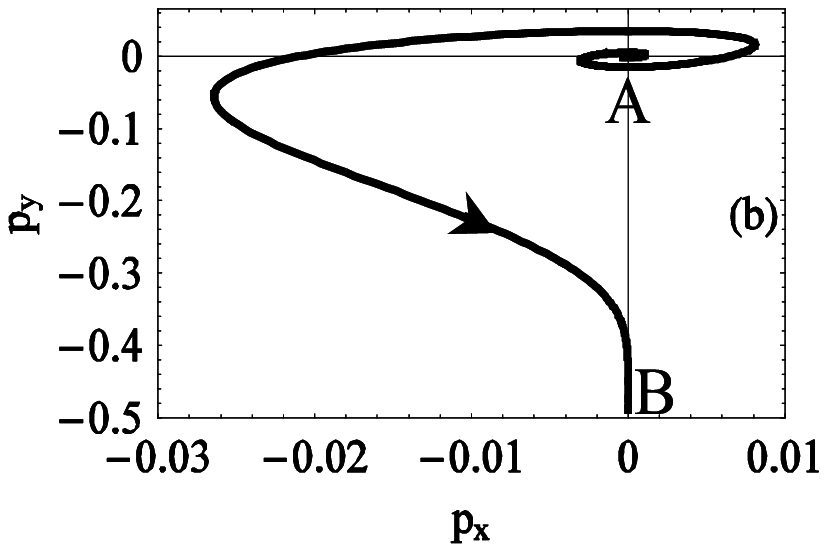} \\
\end{tabular}
\caption{(color online) (a) Projection of the optimal path on the $(x,y)$ plane (thick black line) and the mean-field trajectory ($p_x=p_y=0$) describing an epidemic outbreak (thin red line). (b) Projection of the optimal path on the $(p_x,p_y)$ plane.
$x=S/N-1$, $y=I/N$;  $\,\,\,K=20$ and $\delta=0.5$.} \label{typical1}
\end{figure}

\begin{figure}[ht]
\begin{tabular}{cc}
 \includegraphics[scale=0.52]{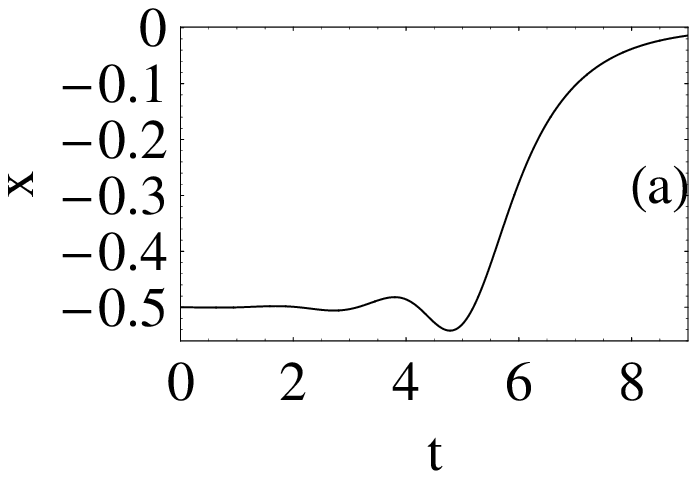} &
 \includegraphics[scale=0.52]{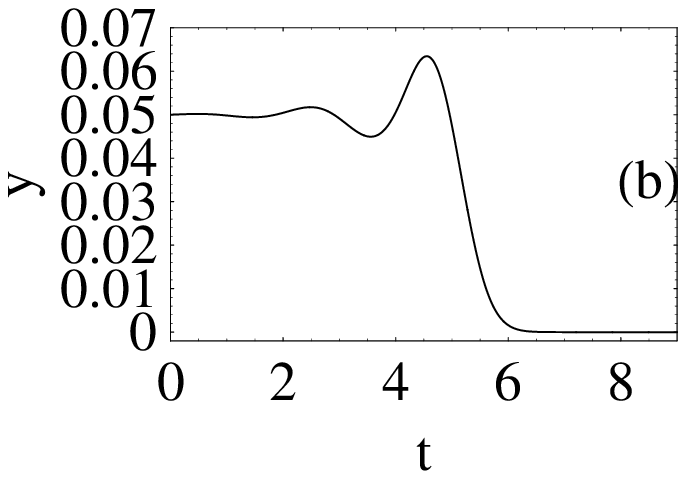}\\
 \includegraphics[scale=0.52]{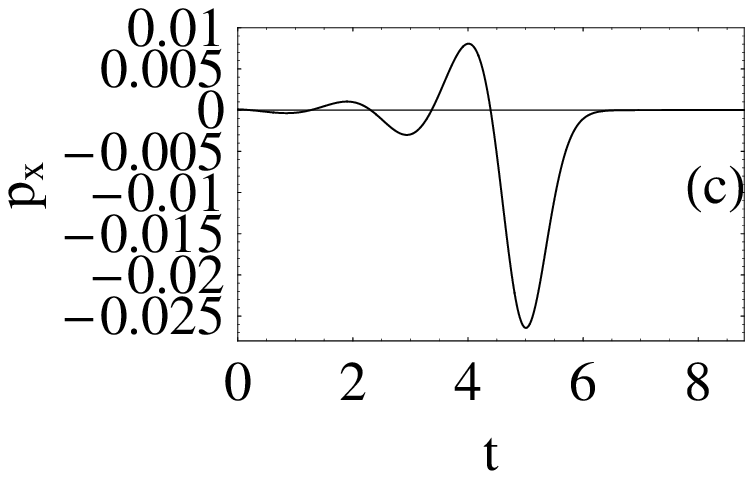} &
 \includegraphics[scale=0.49]{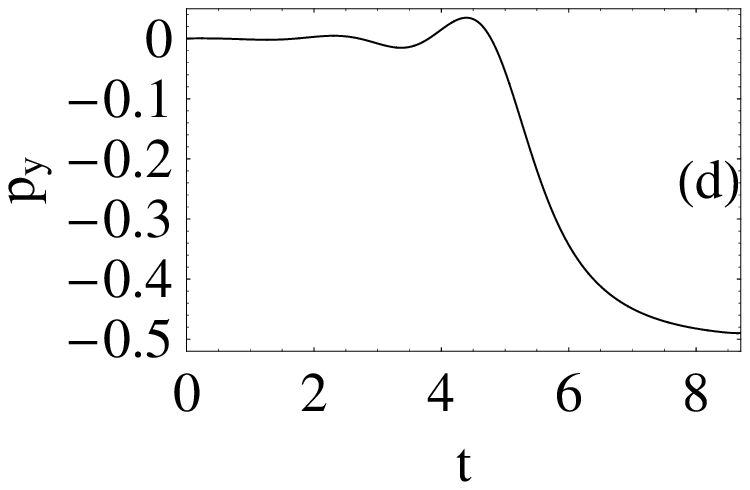}\\
\end{tabular}
\caption{Optimal path for $K=20$ and $\delta=0.5$. Shown are $x=S/N-1$ (a), $y=I/N$ (b), $p_x=p_S-1$ (c) and $p_y=p_I-1$ (d) vs. rescaled time.} \label{typical2}
\end{figure}

\begin{figure}[ht]
\begin{tabular}{cc}
\hspace{-6mm}
  \includegraphics[scale=0.47]{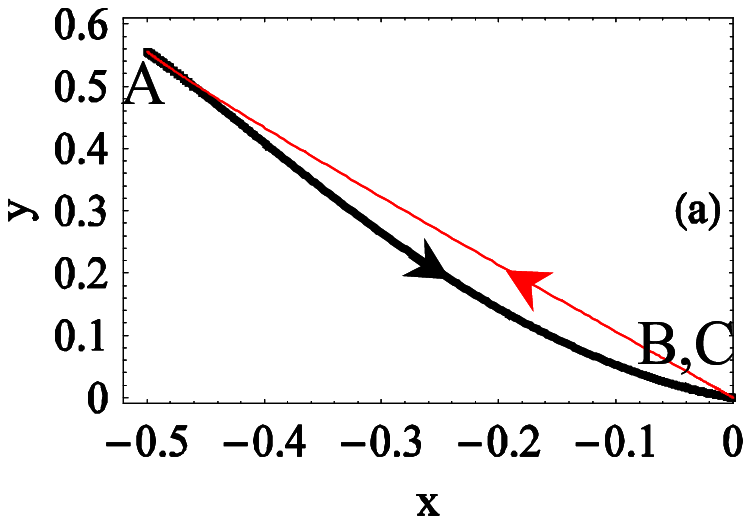} &
  \includegraphics[scale=0.47]{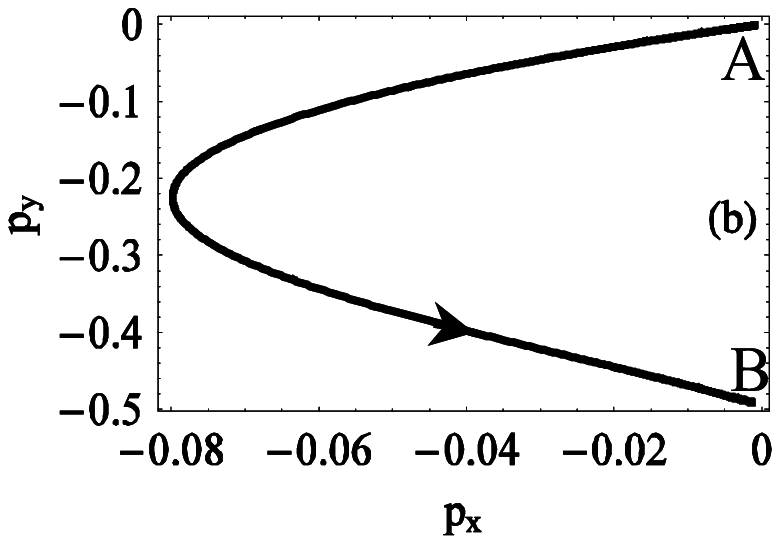} \\
\end{tabular}
\caption{(color online) Same as in Fig.~\ref{typical1} but for $K=1.8$.} \label{typical3}
\end{figure}

\begin{figure}[ht]
\begin{tabular}{cc}
 \includegraphics[scale=0.52]{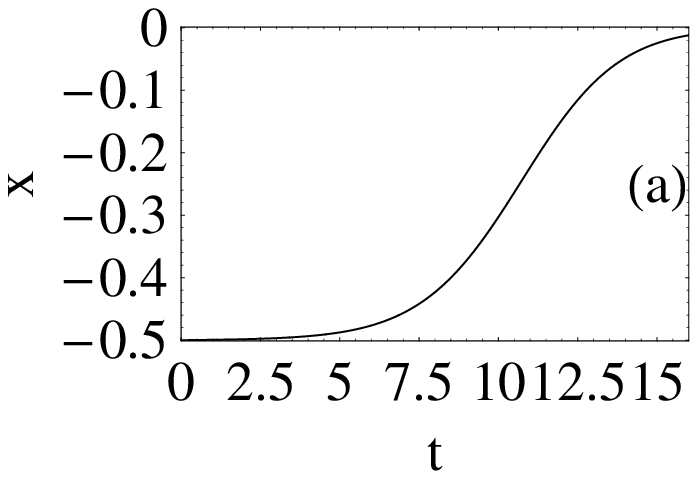} &
 \includegraphics[scale=0.48]{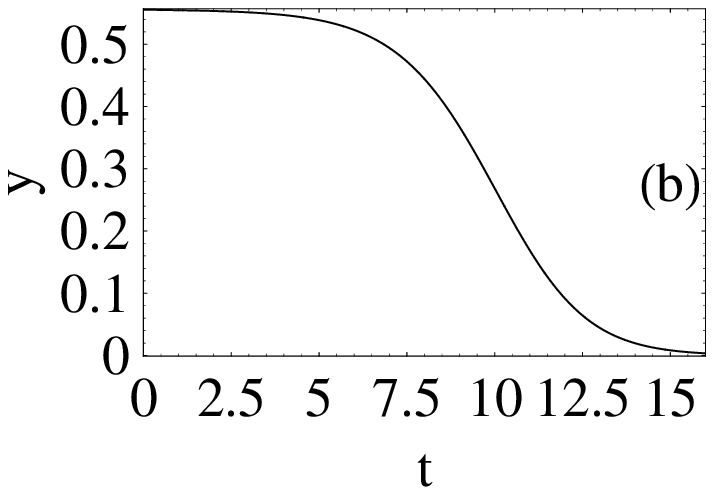}\\
 \includegraphics[scale=0.52]{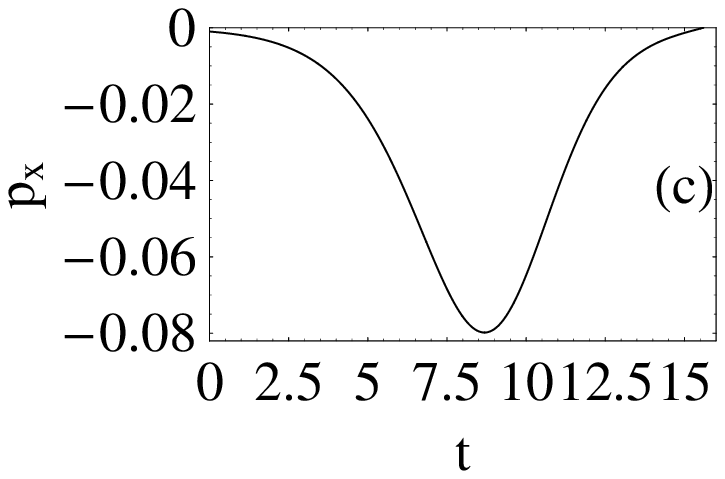} &
 \includegraphics[scale=0.52]{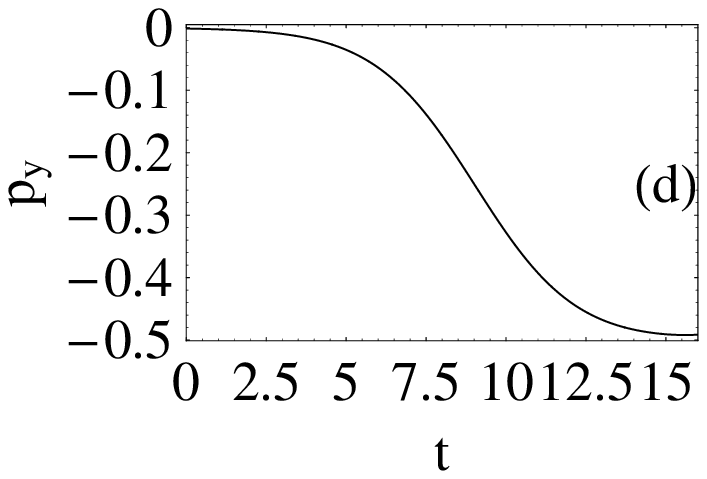}\\
\end{tabular}
\caption{Same as in Fig.~\ref{typical2} but for $K=1.8$.}\label{typical4}
\end{figure}

\textit{Optimal path and action: numerical examples.} We computed the optimal path numerically for different parameters. To find the optimal path one needs to adjust a \textit{single} shooting parameter: the angle between  two unstable eigendirections  of the endemic fixed point $A$.   Two typical examples of  numerically computed optimal paths are shown in Figs.~\ref{typical1} and \ref{typical2} [where $4 K \delta (1-\delta)^2>1$, and the  endemic point is a focus] and in Figs.~\ref{typical3} and \ref{typical4} [where $4K \delta (1-\delta)^2<1$, and the endemic point is a node]. Figures~\ref{typical1}a and \ref{typical3}a show projections of the optimal paths on the $(x,y)$ plane. For comparison, they also show the mean-field trajectories ($p_x=p_y=0$) originating in the vicinity of the no-disease point $x=y=0$, describing an epidemic outbreak and approaching the endemic point. In contrast to equilibrium systems, the optimal path of a large fluctuation is different from the corresponding relaxation path.   Notice that, although for $K=20$ the extinction proceeds along a spiral,
the difference between the two spirals is striking. Figures~\ref{typical1}b and \ref{typical3}b show projections of the optimal paths  on the $(p_x,p_y)$ plane. The optimal paths are presented in more detail in Figs.~\ref{typical2} and \ref{typical4}, where the time dependences of $x$, $y$, $p_x$ and $p_y$ are shown. The rescaled action  along the optimal path in this examples is $\sigma\simeq 6.12 \times 10^{-3}$ for $K=20$ and $\sigma \simeq 0.145$ for $K=1.8$, providing sharp estimates to the logarithm of the corresponding mean extinction times of the disease.

\textit{Optimal path and action: asymptotic theory in the vicinity of the bifurcation point.}
For $K \delta \ll 1$ we can compute the optimal path and the rescaled action $\sigma (K,\delta)$
analytically, by exploiting time scale separation. Let us introduce rescaled  variables: $q_1=x/\delta$, $q_2=yK/\delta$,
$p_1=p_x/\delta^2$, and $p_2=p_y/\delta$. This rescaling is motivated by the values of the coordinates of the fixed points, and reflects the important feature that, at $\delta \ll 1$, the fluctuations in the number of susceptible are much weaker than the fluctuations in the number of infected. Neglecting higher order terms in $\delta$, we arrive at the following approximate equations of motion:
\begin{eqnarray}
\left.\begin{array}{rcl}
\dot{q}_1 &=& -q_1-q_2\,,
\\
\dot{p}_1 &=& p_1 - q_2 p_2\,,\\
\end{array}\right.
\left.\begin{array}{rcl}
\dot{q}_2 &=& K \delta \, q_2(1+q_1+2 p_2)\,,\\
\dot{p}_2 &=& K \delta \,(p_1-p_2-p_2^2-q_1p_2)\,.
\end{array}\right.
\label{4equations}
\end{eqnarray}
The fixed points become $A=[-1,1,0,0]$, $B=[0,0,0,-1]$, and $C=[0,0,0,0]$. For $K\delta\ll 1$ the subsystem $(q_1,p_1)$ is fast, whereas  $(q_2,p_2)$ is slow. On the fast time scale (that is, the time scale $\mu^{-1}$ in the original, dimensional variables) the fast subsystem approaches the state $q_1\simeq -q_2$ and $p_1 \simeq q_2 p_2$ which then slowly evolves according to the equations
\begin{equation}\label{2equations}
    \dot{q}_2 \simeq K \delta \, q_2(1-q_2+2 p_2)\,,\;\;\;\dot{p}_2 \simeq K \delta \,p_2(2q_2 -1-p_2)
\end{equation}
that are Hamiltonian, as they follow from the reduced Hamiltonian $ H_r(q_2,p_2)=K\delta\;q_2 p_2 (1-q_2+p_2)$.
This Hamiltonian  appears in the theory of a class of \textit{single}-species models in the vicinity of a bifurcation point \cite{Kamenev2}.
As $H_r(q_2,p_2)$ is independent of time, it is an integral of motion.  The optimal extinction path goes along the zero-energy trajectory $1-q_2+p_2=0$ \cite{explicit}. Evaluating the action~(\ref{action1}) along this line, we find in the leading order:
${\cal S}_0\simeq \left[N \delta^3/(K\delta)\right]\int_1^0 p_2dq_2= N\delta^2/(2K)$. For the mean extinction time of the disease we obtain
\begin{equation}\label{tau}
 \ln (\tau)/N \simeq \delta^2/(2 K) = [\mu/(2 \beta)]\left(1-\Gamma/\beta\right)^2 \,;
\end{equation}
this asymptote is valid when ${\cal S}_0\gg 1$.

Dykman \textit{et al.} \cite{DykmanSIS} have recently shown that reduced Hamiltonian dynamics of the same type as Eqs.~(\ref{2equations})
holds, close to the bifurcation point, in the endemic SIS model: still another two-population stochastic epidemic model where the infected individuals again become susceptible upon recovery.

In summary, we have developed the eikonal theory for stochastic multi-population birth-death systems. The  theory is especially suitable for analysis of large fluctuations, such as disease extinction. For the SI and SIR models we have found the optimal path to extinction of the disease and the mean extinction time. The optimal path to extinction, including its remarkable oscillatory behavior, is not model-specific.  It should be observable in stochastic simulations of a broad class of models,  and in real data on fade out, of infectious diseases in small communities.

We are very grateful to M. Dykman for advice and for sharing the results of work \cite{DykmanSIS} prior to publication, and to M. Assaf and I. N{\aa}sell for helpful  discussions. B.~M. thanks FTPI of the University of Minnesota for hospitality.  A.~K. was supported by the NSF grant  DMR-0405212  and by the
A.~P.~Sloan foundation; B.~M. was supported by the Israel Science Foundation.


\begin{thebibliography}{99}
\bibitem{spectral} For a single species the equation $\partial_t G=\hat{H}G$
can be analyzed by the spectral theory \cite{Assaf}. This theory, however,  has not been extended to multiple species.
\bibitem{explicit} One can easily compute $q_2 (t) = \left[e^{K \delta (t-t_0)}+1 \right]^{-1}$ and $p_2 (t) = -\left[e^{-K \delta (t-t_0)}+1 \right]^{-1}$, where $t_0=const$.
\end{thebibliography}
\end{document}